\documentclass[nofootinbib,aps,a4paper,letterpaper,superscriptaddress,
twocolumn,times,eqsecnum]{revtex4}
\usepackage{amsmath}
\usepackage{amsfonts}
\usepackage{booktabs}
\usepackage{multirow}
\usepackage{siunitx} 
\usepackage{adjustbox}
\pdfoutput=1
\usepackage{graphicx}
\usepackage{color}
\usepackage{braket}
\usepackage{dcolumn}
\usepackage{bm,url}
\usepackage[linktocpage]{hyperref}
\usepackage{subfigure}
\usepackage{amsfonts}
\usepackage[usenames,dvipsnames,svgnames]{xcolor}  
\usepackage{hyperref}   
\definecolor{oxfordblue}{rgb}{0.0, 0.13, 0.28}
\definecolor{burgundy}{rgb}{0.5, 0.0, 0.13}
\definecolor{darkolivegreen}{rgb}{0.33, 0.42, 0.18}
\definecolor{darkblue}{rgb}{0,0,0.5}
\definecolor{richcarmine}{rgb}{0.84, 0.0, 0.25}
\definecolor{darkblue}{rgb}{0,0,0.5}
\definecolor{bluer}{rgb}{0.00,0.50,0.75}{}
\hypersetup{colorlinks=true, citecolor=red, linkcolor=blue,
 urlcolor = magenta, filecolor=magenta}

\begin{document}
 
 \newcommand\be{\begin{equation}}
  \newcommand\ee{\end{equation}}
 \newcommand\bea{\begin{eqnarray}}
  \newcommand\eea{\end{eqnarray}}
 \newcommand\bseq{\begin{subequations}} 
  \newcommand\eseq{\end{subequations}}
 \newcommand\bcas{\begin{cases}}
  \newcommand\ecas{\end{cases}}
 \newcommand{\p}{\partial}
 \newcommand{\f}{\frac}

\title{Bohmian Quantum Cosmology from the Wheeler-DeWitt Equation}

\author{Spyros Basilakos}
\email{svasil@academyofathens.gr}
\affiliation{Academy of Athens, Research Center for Astronomy $\&$ Applied 
Mathematics, Soranou Efessiou 4, 11-527, Athens, Greece }
\affiliation{National Observatory of Athens, Lofos Nymfon, 11852 Athens, Greece}

\author{Gerasimos Kouniatalis}
\email{gkouniatalis@noa.gr}
 \affiliation{National Observatory of Athens, Lofos Nymfon, 11852 Athens, 
Greece}
\affiliation{Physics Department, National Technical University of Athens,
15780 Zografou Campus,  Athens, Greece}

\author{Emmanuel N. Saridakis} \email{msaridak@noa.gr}
\affiliation{National Observatory of Athens, Lofos Nymfon, 11852 Athens, Greece}
\affiliation{CAS Key Laboratory for Researches in Galaxies and Cosmology, 
School 
of Astronomy and Space Science, University of Science and Technology of China, 
Hefei, Anhui 230026, China}
\affiliation{Departamento de Matem\'{a}ticas, Universidad Cat\'{o}lica del 
Norte, Avda. Angamos 0610, Casilla 1280 Antofagasta, Chile}

\author{Charalampos Tzerefos}
\email{chtzeref@phys.uoa.gr}
\affiliation{National Observatory of Athens, Lofos Nymfon, 11852 Athens, Greece}
\affiliation{Department of Physics, National \& Kapodistrian University of 
Athens, Zografou Campus GR 157 73, Athens, Greece}


\begin{abstract}
We construct a Bohmian quantum cosmological model for a spatially flat
Friedmann-Robertson-Walker universe filled with a single scalar field whose
potential provides a unified description of cold dark matter and dark energy at
the background level. Starting from the Einstein-Hilbert action supplemented
by a scalar field, we derive the minisuperspace Lagrangian and the associated
canonical Hamiltonian formulation. By means of a nontrivial canonical
transformation, the minisuperspace dynamics is mapped into that of a
two-dimensional hyperbolic oscillator with a fixed frequency ratio, rendering
the Wheeler-DeWitt equation exactly solvable by separation of variables.
The resulting Wheeler-DeWitt solutions are expressed in terms of parabolic
cylinder functions and are parametrised by a continuous separation constant,
reflecting the constrained nature of the theory and the absence of a standard
Schrödinger time parameter. Adopting the de~Broglie-Bohm formulation, we derive
deterministic guidance equations in minisuperspace and construct a well-defined
Bohmian Hubble parameter directly in terms of the pilot-wave phase. Finally, we
present a Wheeler-DeWitt-derived toy wave function for which the Bohmian
trajectories and the associated cosmological expansion history can be obtained
analytically, reproducing the late-time $\Lambda$CDM behaviour while exhibiting
quantum modifications at earlier epochs.
\end{abstract}

\maketitle

\section{Introduction}

Quantum cosmology aims at extending quantum principles to the universe as a 
whole, treating spacetime geometry and matter on the same quantum footing. In 
contrast to quantum field theory on a fixed background, the gravitational field 
itself becomes a dynamical quantum variable. In the canonical approach 
pioneered by DeWitt \cite{DeWitt1967}, the dynamics of gravity and matter is 
encoded in the Wheeler-DeWitt (WDW) equation,
which plays the role of a 
Schrödinger equation for the wave function of the universe 
\cite{Wheeler:1968Superspace,Hartle:1983ai,Page1984, 
Kiefer1988,Halliwell:1988Wheeler, 
Falciano2012,Kiefer:2008Cosmology}. 
Additionally, 
foundational ideas related to superspace, quantum creation of the universe, and
inflationary quantum cosmology were further developed   in 
\cite{Vilenkin:1984Creation,
Linde:1984InflationaryCreation}.

Solving the full Wheeler-DeWitt equation, involving infinitely many degrees of 
freedom, is far beyond present analytical and numerical capabilities. A 
standard and physically motivated simplification is therefore necessary to 
restrict 
attention to highly symmetric cosmological models, where only a finite number 
of 
degrees of freedom remain. In particular, in homogeneous and isotropic 
Friedmann-Robertson-Walker 
spacetimes, the gravitational sector description reduces to the scale factor, 
while matter 
can be described by homogeneous fields. This truncation of the 
infinite-dimensional superspace of general relativity to a finite-dimensional 
configuration space is known as minisuperspace 
\cite{Misner:1969hg,HawkingWu,Kuchar:1989tj,Hartle:1983ai}. 
Minisuperspace quantum cosmology has long served as a valuable framework for 
addressing conceptual issues such as the problem of time, boundary conditions 
for the universe, and the emergence of classical cosmological behaviour 
\cite{Vilenkin:1994rn,Kiefer:2005AnnPhysReview,Giulini:1994dx,Gousheh:2000au,
Garattini:2014rwa}.

In this work we consider a spatially flat Friedmann-Robertson-Walker universe 
filled 
with a single scalar field. The scalar field potential is chosen such that, 
near its minimum, it behaves as a massive field superimposed on a cosmological 
constant. The oscillatory massive component can effectively mimic cold dark 
matter, while the constant term drives late-time accelerated expansion, playing 
the role of dark energy. In this way, a single scalar degree of freedom 
provides a unified description of the dark sector at the background level, an 
idea that has attracted considerable interest in both classical and quantum 
cosmology 
\cite{Scherrer2004,Bertacca2007,Arbey:2006it, Kouniatalis:2025qfz, Lim:2010yk, 
Kouniatalis:2025orn,
Piattella:2009kt,Koutsoumbas:2017fxp,ColinPintoNeto:2017MatterBounceDE,
BacalhauPintoNetoVitenti:2018Spectra,   Leon:2022oyy}.

The quantum dynamics of the model is governed by the Wheeler-DeWitt equation 
in minisuperspace. A key element of our analysis is a nontrivial canonical 
transformation that maps the minisuperspace Hamiltonian into the form of a 
two-dimensional hyperbolic oscillator with a fixed relation between its 
frequencies. This reformulation allows the Wheeler-DeWitt equation to be 
solved explicitly by separation of variables. Due to the Hamiltonian constraint 
and the absence of an external Schrödinger time parameter, the standard 
harmonic-oscillator energy quantisation does not apply. Instead of discrete 
levels, the Wheeler-DeWitt equation admits a continuous family of solutions 
labelled by a separation constant, a feature that is often overlooked in the 
minisuperspace literature 
\cite{Kiefer:2005AnnPhysReview,Giulini:1994dx,VargasMoniz:2010zz}.

To extract physical cosmological dynamics from the Wheeler-DeWitt framework, 
an interpretation of the wave function of the universe is required. In our 
analysis we adopt the de Broglie-Bohm (pilot-wave) formulation of quantum 
theory 
\cite{deBroglie1926,Bohm1952,BohmHiley1993,Holland1993,PintoNeto:2004BohmQC,
PintoNeto:2021dBBQCReview,PintoNetoStruyve:2019BohmQGCosmo}. In this approach, 
the 
wave function satisfies the Wheeler-DeWitt equation as usual, but the 
minisuperspace variables follow deterministic trajectories guided by the phase 
of the wave function. The Bohmian formulation is particularly natural in 
quantum cosmology, where no external classical measuring apparatus exists and 
one is interested in the evolution of a single universe rather than in 
ensembles of measurements 
\cite{Shojai2005,PintoNetoFabris:2013dBBQC,Colistete:1997sf}, while 
  cosmological evolutions arising from the quantum potential in
the Bohmian framework have been analyzed within inflationary
minisuperspace models 
 \cite{FalcianoPintoNetoSantini:2007InflationaryNS}.

The novelty of the present analysis lies in the combination of three elements: 
(i) an exactly solvable minisuperspace model with a unified dark sector scalar 
field, (ii) a transparent reformulation of the Wheeler-DeWitt equation as a 
hyperbolic oscillator admitting a continuous spectrum, and (iii) a fully 
explicit Bohmian construction yielding analytic trajectories and a 
well-defined Bohmian Hubble parameter. In particular, we present a 
Wheeler-DeWitt-derived toy wave function for which the Bohmian trajectories 
and the resulting cosmological expansion history can be obtained analytically, 
allowing a direct comparison with the classical $\Lambda$CDM background.

The paper is organised as follows. In Section \ref{Themodel} we present the 
classical minisuperspace model in a flat Friedmann-Robertson-Walker geometry 
and discuss 
the physical interpretation of the scalar field and its potential. 
Section~\ref{WheelerDeWitt}    is devoted to the canonical formulation and the 
explicit solution of the Wheeler-DeWitt equation. In Section 
\ref{BohmianCosmology} we construct the Bohmian quantum cosmology, derive the 
guidance equations, and introduce the Bohmian Hubble parameter. Finally, in 
Section \ref{Conclusions}  we summarise our results and outline possible 
extensions.

\section{Quantum cosmology in the FRW geometry}
\label{Themodel}

In this section we briefly review the classical minisuperspace model in 
a flat
Friedmann-Robertson-Walker (FRW) geometry, and we discuss the
physical interpretation of the scalar field and the unified dark matter-dark 
energy potential. 
Throughout the work we adopt natural units 
$
c = 1,   \hbar = 1, 8\pi G = 1,
$
where $c$ denotes the speed of light, $\hbar$ the reduced Planck constant, and 
$G$ Newton's constant.

We consider a spatially homogeneous and isotropic universe described by the 
spatially flat FRW metric
\begin{equation}
ds^{2} = -dt^{2} + \alpha(t)^{2} \left( dx^{2} + dy^{2} + dz^{2} \right),
\label{eq:rwmetric}
\end{equation}
where $t$ denotes cosmic time and $\alpha(t)$ is the scale factor. The scale 
factor encodes the relative expansion of comoving spatial distances, while the 
Hubble 
parameter is defined as
\begin{equation}
H(t) \equiv \frac{\dot{\alpha}(t)}{\alpha(t)},
\label{eq:hubble}
\end{equation}
and characterizes the expansion rate of the universe at a given cosmic time.
Finally, we recall that the general Einstein field equations take the 
compact form
\begin{equation}
G_{\mu\nu} \equiv R_{\mu\nu} - \frac{1}{2} g_{\mu\nu} R = T_{\mu\nu},
\label{eq:einstein}
\end{equation}
with $R_{\mu\nu}$ the Ricci tensor, $R$ the Ricci scalar, $g_{\mu\nu}$ the 
spacetime metric, and $T_{\mu\nu}$ the energy-momentum tensor.

\subsection{Scalar field matter content}

As a matter source we assume a single real scalar field $\phi(t)$, homogeneous 
on the spatial hypersurfaces. Its action is given by
\begin{equation}
S_{\phi} = \int d^{4}x \, \sqrt{-g} 
\left[ \frac{1}{2} g^{\mu\nu} \partial_{\mu}\phi \, \partial_{\nu}\phi - 
U(\phi) \right],
\label{eq:scalaraction}
\end{equation}
where $U(\phi)$ denotes the scalar field potential. The corresponding 
energy-momentum tensor reads
\begin{equation}
T_{\mu\nu}(\phi) = \partial_{\mu}\phi \, \partial_{\nu}\phi 
 - g_{\mu\nu} \left( \frac{1}{2} g^{\rho\sigma} \partial_{\rho}\phi \, 
\partial_{\sigma}\phi - U(\phi) \right).
\label{eq:scalarTmunu}
\end{equation}

For a homogeneous configuration $\phi=\phi(t)$ in the metric 
\eqref{eq:rwmetric}, the energy density and isotropic pressure are
\begin{equation}
\rho_{\phi} = T^{0}{}_{0} = \frac{1}{2} \dot{\phi}^{2} + U(\phi),
\label{eq:rho_phi}
\end{equation}
\begin{equation}
p_{\phi} = T^{i}{}_{i} = \frac{1}{2} \dot{\phi}^{2} - U(\phi),
\quad \text{(no sum on $i$)}.
\label{eq:p_phi}
\end{equation}
The kinetic contribution corresponds to a stiff equation of state $p=\rho$, 
while the potential energy behaves as an effective cosmological constant with 
$p=-\rho$. The relative dominance of these contributions determines the 
effective cosmological behaviour of the scalar field during the evolution of 
the universe.

\subsection{Minisuperspace Lagrangian and Friedmann equations}

Substituting the metric \eqref{eq:rwmetric} and the homogeneous scalar field 
into the Einstein-Hilbert action supplemented by \eqref{eq:scalaraction}, and 
integrating over the comoving spatial volume, one obtains an effective 
one-dimensional minisuperspace Lagrangian for the variables $\alpha(t)$ and 
$\phi(t)$ 
\cite{Paliathanasis:2014iva,Paliathanasis:2015aos,Dimakis:2016mip}:
\begin{equation}
L(\alpha, \dot{\alpha}, \phi, \dot{\phi}) 
= -3 \alpha \dot{\alpha}^{2} 
+ \frac{1}{2} \alpha^{3} \left( \dot{\phi}^{2} - U(\phi) \right),
\label{eq:minilagrangian}
\end{equation}
where an irrelevant overall constant factor has been omitted. This Lagrangian 
describes the coupled dynamics of the scale factor and the homogeneous scalar 
field. The negative sign of the gravitational kinetic term reflects the 
indefinite nature of the minisuperspace metric, a generic feature of the 
canonical formulation of general relativity.

Varying the action $\int L\,dt$ with respect to $\alpha(t)$ and $\phi(t)$ 
yields the Euler-Lagrange equations, which are equivalent to the Friedmann 
equations and the Klein-Gordon equation:
\begin{equation}
3 \left( \frac{\dot{\alpha}}{\alpha} \right)^{2} 
= \frac{1}{2} \dot{\phi}^{2} + U(\phi),
\label{eq:friedmann1}
\end{equation}
\begin{equation}
2 \frac{\ddot{\alpha}}{\alpha} 
+ \left( \frac{\dot{\alpha}}{\alpha} \right)^{2} 
= -\frac{1}{2} \dot{\phi}^{2} + U(\phi),
\label{eq:friedmann2}
\end{equation}
\begin{equation}
\ddot{\phi} + 3 \frac{\dot{\alpha}}{\alpha} \dot{\phi} 
+ \frac{dU}{d\phi} = 0.
\label{eq:kg}
\end{equation}
Equation \eqref{eq:friedmann1} represents the Hamiltonian constraint, relating 
the expansion rate to the scalar field energy density. Moreover, equation 
\eqref{eq:friedmann2} governs the acceleration of the expansion, while 
\eqref{eq:kg} describes the scalar field dynamics in an expanding background, 
with the Hubble parameter acting as a friction term.

\subsection{Unified dark matter-dark energy potential}

In order to realise a unified description of dark matter and dark energy, we 
adopt the scalar field potential
\begin{equation}
U(\phi) = \frac{\Lambda}{2} \left( \cosh^{2}(c\phi) + 1 \right),
\label{eq:potential}
\end{equation}
where $\Lambda>0$ has the dimensions of an energy density and $c$ is a constant 
with dimensions of inverse field. The potential admits a minimum at $\phi=0$, 
around which its Taylor expansion reads
\begin{equation}
U(\phi) = \Lambda + \frac{\Lambda c^{2}}{4} \, \phi^{2} 
+ \mathcal{O}(\phi^{4}).
\label{eq:Uexpansion}
\end{equation}
The constant term acts as an effective cosmological constant, while the 
quadratic contribution corresponds to a massive scalar field with mass
\begin{equation}
m^{2} = \frac{\Lambda c^{2}}{2}.
\end{equation}

When the scalar field undergoes oscillations in the quadratic region of the 
potential, its averaged pressure is negligible compared to its averaged energy 
density, and it therefore behaves as cold dark matter. At late times, as the 
field settles near the minimum, the constant term dominates and the 
cosmological dynamics approaches a dark-energy-dominated phase. Consequently, 
the same scalar field can effectively describe both cold dark matter and dark 
energy at different stages of cosmic evolution.

\section{Wheeler-DeWitt quantisation and exact solutions}
\label{WheelerDeWitt}

We now turn to the canonical quantisation of the minisuperspace dynamics and 
the construction of the Wheeler-DeWitt equation. In particular, a suitable 
choice of variables recasts the Hamiltonian into a form amenable to exact 
solution, revealing the underlying hyperbolic-oscillator structure of the 
quantum cosmological system.

\subsection{Canonical formulation and hyperbolic oscillator}

For the purpose of quantisation it is convenient to pass from the Lagrangian 
formulation \eqref{eq:minilagrangian} to the Hamiltonian formulation. The 
canonical momenta conjugate to $\alpha$ and $\phi$ are given by
\begin{equation}
\pi_{\alpha} \equiv \frac{\partial L}{\partial \dot{\alpha}} = -6 \alpha 
\dot{\alpha},
\label{eq:pi_alpha}
\end{equation}
\begin{equation}
\pi_{\phi} \equiv \frac{\partial L}{\partial \dot{\phi}} = \alpha^{3} 
\dot{\phi}.
\label{eq:pi_phi}
\end{equation}
The canonical Hamiltonian is obtained via the Legendre transform
\begin{equation}
H(\alpha,\pi_{\alpha},\phi,\pi_{\phi})
= \pi_{\alpha} \dot{\alpha} + \pi_{\phi} \dot{\phi} - L.
\end{equation}
Expressing the velocities in terms of the canonical momenta using
Eqs.~\eqref{eq:pi_alpha}-\eqref{eq:pi_phi}, one finds
\begin{equation}
H = -\frac{1}{12\alpha} \pi_{\alpha}^{2}
+ \frac{1}{2\alpha^{3}} \pi_{\phi}^{2}
+ \frac{1}{2} \alpha^{3} U(\phi).
\label{eq:hamiltonian_alpha_phi}
\end{equation}

As it is known, the time reparametrisation invariance of general relativity 
implies that the 
Hamiltonian is a constraint rather than a generator of physical time evolution. 
Consequently, the classical dynamics is restricted by
\begin{equation}
H \equiv 0,
\label{eq:hamconstraint}
\end{equation}
which is equivalent to the Friedmann equation \eqref{eq:friedmann1}. Upon 
quantisation, this constraint becomes the Wheeler-DeWitt equation.

In order to simplify the structure of the Hamiltonian, we introduce a canonical 
transformation from $(\alpha,\phi)$ to new variables $(x,y)$ defined as
\begin{equation}
x = A \, \alpha^{3/2} \sinh(c\phi), \qquad
y = A \, \alpha^{3/2} \cosh(c\phi),
\label{eq:xydef}
\end{equation}
where $A$ is a positive constant to be fixed and $c$ coincides with the 
parameter appearing in the potential \eqref{eq:potential}. Combining these 
definitions yields
\begin{equation}
y^{2} - x^{2} = A^{2} \alpha^{3},
\end{equation}
from which the scale factor can be expressed as
\begin{equation}
\alpha(x,y) = \left( \frac{y^{2} - x^{2}}{A^{2}} \right)^{1/3}.
\label{eq:alpha_xy}
\end{equation}
Similarly, the scalar field is given by
\begin{equation}
\tanh(c\phi) = \frac{x}{y}
\quad \Rightarrow \quad
\phi(x,y) = \frac{1}{c}
\operatorname{arctanh}\!\left(\frac{x}{y}\right).
\label{eq:phi_xy}
\end{equation}
The variables $(x,y)$ thus provide an alternative parametrisation of 
minisuperspace, in which the geometric and matter degrees of freedom are 
nonlinearly mixed.

For a specific choice of the parameters $A$ and $c$, the minisuperspace 
Lagrangian takes the   simple form
\begin{equation}
L(x,\dot{x},y,\dot{y})
= \frac{1}{2} \left( \dot{x}^{2} + \omega_{1}^{2} x^{2} \right)
 - \frac{1}{2} \left( \dot{y}^{2} + \omega_{2}^{2} y^{2} \right),
\label{eq:lagrangian_xy}
\end{equation}
with
\begin{equation}
\omega_{1}^{2} = \frac{\Lambda}{A^{2}}, \qquad
\omega_{2}^{2} = \frac{2\Lambda}{A^{2}}.
\label{eq:omegas}
\end{equation}
Explicitly, one finds
\begin{equation}
A^{2} = \frac{8}{3}, \qquad
c^{2} = \frac{3}{8},
\label{eq:Ac_values}
\end{equation}
implying the fixed frequency relation
\begin{equation}
\omega_{2} = \sqrt{2}\,\omega_{1}.
\label{eq:omega_relation}
\end{equation}
Due to the opposite signs of the kinetic terms, the system corresponds to a 
two-dimensional hyperbolic oscillator.

The canonical momenta conjugate to $x$ and $y$ are
\begin{equation}
\pi_{x} = \dot{x}, \qquad
\pi_{y} = -\dot{y},
\label{eq:pi_xy}
\end{equation}
and the corresponding Hamiltonian reads
\begin{equation}
H(x,\pi_{x},y,\pi_{y})
= \frac{1}{2} \left( \pi_{x}^{2} - \omega_{1}^{2} x^{2} \right)
 - \frac{1}{2} \left( \pi_{y}^{2} + \omega_{2}^{2} y^{2} \right).
\label{eq:ham_xy}
\end{equation}
The Hamiltonian constraint $H\equiv0$ must again be imposed. In this 
representation, the minisuperspace dynamics is equivalent to the difference of 
two harmonic oscillators with a fixed frequency ratio, a structure that will 
prove particularly convenient for quantisation.

\subsection{Quantisation and Wheeler-DeWitt equation}

The canonical quantisation proceeds by promoting the canonical variables to 
operators acting on the minisuperspace wave function $\Psi(x,y)$. The momenta 
are represented as differential operators,
\begin{equation}
\pi_{x} \rightarrow \hat{\pi}_{x} = - i \frac{\partial}{\partial x},
\qquad
\pi_{y} \rightarrow \hat{\pi}_{y} = - i \frac{\partial}{\partial y},
\label{eq:quantisation}
\end{equation}
leading to the Wheeler-DeWitt operator
\begin{equation}
\hat{H}
= -\frac{1}{2} \frac{\partial^{2}}{\partial x^{2}}
 - \frac{1}{2} \omega_{1}^{2} x^{2}
 + \frac{1}{2} \frac{\partial^{2}}{\partial y^{2}}
 + \frac{1}{2} \omega_{2}^{2} y^{2}.
\label{eq:wdw_operator}
\end{equation}
The Wheeler-DeWitt equation is the operator implementation of the Hamiltonian 
constraint,
\begin{equation}
\hat{H}\Psi(x,y)=0.
\label{eq:wdw}
\end{equation}
Explicitly, it takes the form
\begin{equation}
\left(
- \frac{\partial^{2}}{\partial x^{2}}
+ \frac{\partial^{2}}{\partial y^{2}}
- \omega_{1}^{2} x^{2}
+ \omega_{2}^{2} y^{2}
\right)\Psi(x,y)=0.
\label{eq:wdw_explicit}
\end{equation}
This equation has the structure of a Klein-Gordon equation in a 
two-dimensional minisuperspace with coordinates $(x,y)$ and an effective 
potential
\begin{equation}
V(x,y) = -\omega_{1}^{2} x^{2} + \omega_{2}^{2} y^{2}.
\end{equation}
Note that the underlying minisuperspace metric has Lorentzian signature 
$(-,+)$, 
reflecting the negative kinetic contribution of the gravitational sector.
We note that different choices of operator ordering in the Wheeler-DeWitt
quantisation may affect the resulting quantum dynamics and, in particular,
the resolution of cosmological singularities (see 
 \cite{DemaerelStruyve:2019Orderings} for a detailed analysis).

\subsection{Separation of variables and parabolic cylinder functions}

We seek solutions of Eq.~\eqref{eq:wdw_explicit} by separation of variables, 
namely
\begin{equation}
\Psi(x,y)=X(x)Y(y).
\label{eq:sep_ansatz}
\end{equation}
Substitution into the Wheeler-DeWitt equation yields two ordinary differential 
equations,
\begin{equation}
X''(x)+\left(\omega_{1}^{2}x^{2}-2E\right)X(x)=0,
\label{eq:x_eq}
\end{equation}
\begin{equation}
Y''(y)+\left(\omega_{2}^{2}y^{2}-2E\right)Y(y)=0,
\label{eq:y_eq}
\end{equation}
where $E$ is a separation constant. Although these equations resemble 
Schrödinger equations for harmonic oscillators, the parameter $E$ does not 
correspond to a physical energy eigenvalue, but arises from the Hamiltonian 
constraint.

The general equation
\begin{equation}
f''(q)+\left(\omega^{2}q^{2}-2E\right)f(q)=0
\label{eq:general_osc}
\end{equation}
admits solutions in terms of parabolic cylinder functions. Introducing the 
dimensionless variable $z=\sqrt{\omega}\,q$ and performing a suitable complex 
rescaling, the independent solutions can be expressed as
\begin{equation}
f(q)
= C_{1}\,D_{\nu}\!\left(e^{i\pi/4}\sqrt{2\omega}\,q\right)
+ C_{2}\,D_{-\nu-1}\!\left(i e^{i\pi/4}\sqrt{2\omega}\,q\right),
\end{equation}
with
\begin{equation}
\nu=-\frac{1}{2}+\frac{iE}{\omega}.
\end{equation}

The symbol $D_\nu(\cdot)$ denotes the
standard parabolic cylinder function (also known as Weber's function), i.e. a
canonical solution of Weber's differential equation \cite{Olver:2010:NHMF}
\begin{equation}
\frac{d^2 f}{dz^2}+\left(\nu+\frac{1}{2}-\frac{z^2}{4}\right)f=0.
\end{equation}

In ordinary quantum mechanics, the requirement of square integrability selects 
a discrete energy spectrum. In the present minisuperspace context, however, the 
Wheeler-DeWitt equation is a constraint, no external time parameter exists, 
and the natural inner product is indefinite. Consequently, the standard 
arguments leading to discrete harmonic-oscillator levels do not apply, and 
there is no fundamental reason to restrict $E$ to a discrete set of values. 
Instead, $E$ is naturally treated as a continuous parameter. 
 The appearance of a continuous spectrum is therefore not a peculiarity 
of the present model, but a generic consequence of the Hamiltonian constraint 
and the Klein-Gordon-type structure of the Wheeler-DeWitt equation. 
Moreover, the continuous nature of $E$ plays a crucial role in defining 
admissible Bohmian trajectories and the resulting quantum-cosmological 
dynamics.  

Accordingly, the separated solutions can be written as
\begin{align}
X(x;E) =& A_{1}(E) D_{\nu_{1}(E)}\!\left( e^{i\pi/4} \sqrt{2\omega_{1}}\, x 
\right) \nonumber \\
 &+ A_{2}(E) D_{-\nu_{1}(E)-1}\!\left( i e^{i\pi/4} \sqrt{2\omega_{1}}\, x 
\right),
\end{align}
\begin{align}
Y(y;E) =& B_{1}(E) D_{\nu_{2}(E)}\!\left( e^{i\pi/4} \sqrt{2\omega_{2}}\, y 
\right) \nonumber \\
 &+ B_{2}(E) D_{-\nu_{2}(E)-1}\!\left( i e^{i\pi/4} \sqrt{2\omega_{2}}\, y 
\right),
\end{align}
where
\begin{equation}
\nu_{1}(E)=-\frac{1}{2}+\frac{iE}{\omega_{1}},
\qquad
\nu_{2}(E)=-\frac{1}{2}+\frac{iE}{\omega_{2}}.
\end{equation}
A separable Wheeler-DeWitt solution is then
\begin{equation}
\Psi_{E}(x,y)=X(x;E)Y(y;E),
\end{equation}
and the general minisuperspace wave function may be constructed as a 
superposition,
\begin{equation}
\Psi(x,y)=\int_{\mathcal{C}} dE\,
A(E)\,X(x;E)Y(y;E),
\label{eq:generalPsi}
\end{equation}
where $\mathcal{C}$ denotes a contour in the complex $E$-plane and $A(E)$ is a 
weight function fixed by boundary conditions or physical regularity 
requirements. In this way, the quantum state of the universe is built from 
elementary modes labelled by a continuous separation constant.

\section{The Bohmian Quantum Cosmology}
\label{BohmianCosmology}

In order to extract a physically meaningful cosmological dynamics from the
Wheeler-DeWitt framework, an interpretation of the wave function of the
universe is required. In this section we adopt the de~Broglie-Bohm
(pilot-wave) formulation of quantum theory, in which the wave function
governs the evolution of the minisuperspace variables through deterministic
guidance equations. This approach is particularly suited to quantum
cosmology, where no external classical observer exists and one is interested
in the description of individual cosmological histories rather than
measurement outcomes. For a recent and comprehensive discussion of the problem 
of time and related
interpretational issues in quantum cosmology, see 
Ref. \cite{KieferPeter:2022TimeQC}.

\subsection{Polar decomposition and guidance equations}

In the de Broglie-Bohm formulation of quantum theory, the wave function is 
supplemented by actual trajectories for the configuration variables. In the 
present minisuperspace model, the configuration space is two-dimensional, with 
coordinates $(x,y)$. The wave function $\Psi(x,y)$ is written in polar form as
\begin{equation}
\Psi(x,y) = R(x,y)\, e^{i S(x,y)},
\label{eq:polar}
\end{equation}
where $R(x,y)\geq0$ denotes the amplitude and $S(x,y)$ the real phase. Given an 
explicit expression for $\Psi$, one may compute
\begin{equation}
R(x,y)=|\Psi(x,y)|=
\sqrt{[\Re\Psi(x,y)]^{2}+[\Im\Psi(x,y)]^{2}},
\end{equation}
\begin{equation}
S(x,y)=\arg\Psi(x,y),
\end{equation}
up to an irrelevant additive multiple of $2\pi$.

Although the Wheeler--DeWitt equation (\ref{eq:wdw}) is a Hamiltonian constraint and thus
does not contain an external Schr\"odinger time, in the de~Broglie--Bohm approach the wave
function $\Psi$ is not interpreted as an evolving state but rather as a time-independent
guiding field on minisuperspace.
Writing $\Psi=R\,e^{iS}$, the phase $S(q)$ defines the momenta through the Hamilton--Jacobi-type
identification
\begin{equation}
p_A=\partial_A S\,,
\label{eq:HJ_identification}
\end{equation}
without introducing any additional $\partial S/\partial t$ term in the Wheeler--DeWitt equation.
To obtain a parametrized evolution for the actual configuration $q^A(t)$ one combines
(\ref{eq:HJ_identification}) with the classical relation between momenta and velocities,
\begin{equation}
p_A=\frac{1}{N}\,G_{AB}(q)\,\dot q^{\,B}\,,
\label{eq:momentum_velocity_general}
\end{equation}
where $G_{AB}$ is the minisuperspace metric and $N(t)$ is the lapse. Fixing the lapse corresponds
to a choice of time parametrization (gauge). In our FRW setting we adopt the cosmic/proper-time
gauge $N=1$, so that (\ref{eq:momentum_velocity_general}) reproduces the guidance equations
(\ref{eq:guidance_momenta})--(\ref{xy}). Hence the wave function remains timeless, while the minisuperspace
variables (and therefore the reconstructed scale factor) evolve along Bohmian trajectories
parameterized by the chosen FRW time coordinate.

The Bohmian guidance equations relate the canonical momenta to gradients of the 
phase,
\begin{equation}
\pi_{x}=\frac{\partial S}{\partial x}, \qquad
\pi_{y}=\frac{\partial S}{\partial y},
\label{eq:guidance_momenta}
\end{equation}
where the sign convention is consistent with the minisuperspace Lagrangian 
\eqref{eq:lagrangian_xy}. Using the classical Hamilton equations,
\begin{equation}\label{xy}
\dot{x}=\pi_{x}, \qquad \dot{y}=-\pi_{y},
\end{equation}
one obtains the Bohmian equations of motion for the minisuperspace 
configuration $(x(t),y(t))$,
\begin{equation}
\dot{x}(t)=\left.\frac{\partial S}{\partial x}\right|_{(x,y)=(x(t),y(t))},
\label{eq:bohm_x}
\end{equation}
\begin{equation}
\dot{y}(t)=-\left.\frac{\partial S}{\partial y}\right|_{(x,y)=(x(t),y(t))}.
\label{eq:bohm_y}
\end{equation}
The overdot denotes differentiation with respect to the chosen time parameter, 
which we identify with the cosmic time inherited from the classical model. 
Note that the sign in $\dot{y}=-\pi_{y}$ follows from the indefinite kinetic
structure of the minisuperspace Hamiltonian \eqref{eq:ham_xy}; consistently,
the canonical momentum satisfies $\pi_{y}=\partial L/\partial\dot{y}=-\dot{y}$.
Equations \eqref{eq:bohm_x}-\eqref{eq:bohm_y} show that the phase of the wave 
function acts as a generating function for the quantum trajectories in 
minisuperspace.

\subsection{Quantum Hamilton-Jacobi equation and quantum potential}

Substituting the polar decomposition \eqref{eq:polar} into the Wheeler-DeWitt 
equation \eqref{eq:wdw_explicit} and separating real and imaginary parts yields 
two coupled real equations. One has the form of a continuity equation for the 
probability current in minisuperspace, while the other is a quantum 
Hamilton-Jacobi equation,
\begin{equation}
\left(\frac{\partial S}{\partial x}\right)^{2}
-\left(\frac{\partial S}{\partial y}\right)^{2}
-\omega_{1}^{2}x^{2}
+\omega_{2}^{2}y^{2}
+Q(x,y)=0,
\label{eq:qHJ}
\end{equation}
where the quantum potential is defined as
\begin{equation}
Q(x,y)\equiv
-\frac{1}{R}\left(
\frac{\partial^{2}R}{\partial x^{2}}
-\frac{\partial^{2}R}{\partial y^{2}}
\right).
\label{eq:Q}
\end{equation}
Equation \eqref{eq:qHJ} closely resembles the classical Hamilton-Jacobi 
equation associated with the Hamiltonian \eqref{eq:ham_xy}, with the additional 
term $Q(x,y)$ encoding genuine quantum effects. When the quantum potential is 
negligible, the phase $S$ approximately satisfies the classical 
Hamilton-Jacobi equation and the Bohmian trajectories approach the classical 
solutions. On the other hand, when $Q(x,y)$ is non-negligible, quantum effects 
can significantly 
modify the minisuperspace dynamics.

The quantum potential depends solely on the amplitude $R(x,y)$ of the wave 
function. Consequently, even though the classical minisuperspace potential
$V(x,y)=-\omega_{1}^{2}x^{2}+\omega_{2}^{2}y^{2}$
is smooth and simple, interference effects in $R(x,y)$ can generate a highly 
nontrivial quantum potential landscape. This provides the mechanism through 
which the pilot wave influences the evolution of the universe in minisuperspace.

\subsection{Bohmian Hubble parameter and physical variables}

To connect the Bohmian minisuperspace dynamics with physical cosmological 
observables, it is useful to express the evolution in terms of the scale 
factor. Along a Bohmian trajectory $(x(t),y(t))$, the scale factor is given by
\begin{equation}
\alpha(t)=\left(\frac{y(t)^{2}-x(t)^{2}}{A^{2}}\right)^{1/3}.
\label{eq:alpha_t}
\end{equation}
Differentiation with respect to time yields
\begin{equation}
\dot{\alpha}(t)=
\frac{2}{3A^{2}}
\left(\frac{y^{2}-x^{2}}{A^{2}}\right)^{-2/3}
\left(y\dot{y}-x\dot{x}\right),
\end{equation}
and thus the Bohmian Hubble parameter is  written as
\begin{equation}
H_{\mathrm{Bohm}}(t)\equiv
\frac{\dot{\alpha}(t)}{\alpha(t)}=
\frac{2}{3}
\frac{x(t)\dot{x}(t)-y(t)\dot{y}(t)}
{y(t)^{2}-x(t)^{2}}.
\label{eq:HBohm_xy}
\end{equation}
Using the guidance equations \eqref{eq:bohm_x}-\eqref{eq:bohm_y}, this can be 
written directly in terms of the phase of the wave function as
\begin{equation}
H_{\mathrm{Bohm}}(t)=
\frac{2}{3}
\frac{x(t)\,\partial S/\partial x
+y(t)\,\partial S/\partial y}
{y(t)^{2}-x(t)^{2}}
\Bigg|_{(x,y)=(x(t),y(t))}.
\label{eq:HBohm_S}
\end{equation}
Given an explicit wave function $\Psi(x,y)$, one can therefore compute the 
Bohmian trajectories, reconstruct the scale factor, and obtain a well-defined 
Bohmian expansion rate. This provides a concrete link between the 
Wheeler-DeWitt wave function and an effective cosmological evolution.

\subsection{A Wheeler-DeWitt-derived toy wave function and analytic 
trajectories}

To illustrate the formalism, we consider a simple toy model for which the 
Bohmian dynamics can be obtained analytically. We stress that the purpose of 
the 
toy wave function introduced here is not to provide a full phenomenological 
fit, 
but to demonstrate in a fully analytic manner how Bohmian trajectories and a 
cosmological expansion history can emerge from the Wheeler-DeWitt framework. 
Starting from the general solution \eqref{eq:generalPsi}, we choose a spectral 
weight sharply peaked around a value $E_{\star}$,
\begin{equation}
A(E)=\frac{1}{\sqrt{2\pi}\,\sigma}
\exp\!\left[-\frac{(E-E_{\star})^{2}}{2\sigma^{2}}\right],
\qquad
\sigma\ll|E_{\star}|.
\end{equation}
In the limit $\sigma\to0$, the wave function is dominated by a single separable 
mode,
\begin{equation}
\Psi(x,y)\simeq
\Psi_{\star}(x,y)\equiv
X(x;E_{\star})Y(y;E_{\star}).
\end{equation}

In the semiclassical regime $|x|,|y|\gg1$, the parabolic cylinder functions 
admit a WKB approximation. One finds
\begin{equation}
X(x;E_{\star})\simeq
\frac{C_{x}}{\sqrt{p_{x}(x)}}
\exp\!\left[iS_{x}(x)\right],
\ \ \ 
p_{x}(x)=\sqrt{\omega_{1}^{2}x^{2}+2E_{\star}},
\end{equation}
and similarly
\begin{equation}
Y(y;E_{\star})\simeq
\frac{C_{y}}{\sqrt{p_{y}(y)}}
\exp\!\left[iS_{y}(y)\right],
\ \ \ 
p_{y}(y)=\sqrt{\omega_{2}^{2}y^{2}+2E_{\star}}.
\end{equation}
Additionally, the phases are given by
\begin{equation}
S_{x}(x)=\int^{x}dx'\,p_{x}(x'),\qquad
S_{y}(y)=\int^{y}dy'\,p_{y}(y'),
\end{equation}
 and evaluating the integrals yields
\begin{equation}
S_{x}(x)=\frac{1}{2}xp_{x}(x)
+\frac{E_{\star}}{\omega_{1}}
\operatorname{arsinh}\!\left(
\frac{\omega_{1}x}{\sqrt{2E_{\star}}}
\right),
\end{equation}
\begin{equation}
S_{y}(y)=\frac{1}{2}yp_{y}(y)
+\frac{E_{\star}}{\omega_{2}}
\operatorname{arsinh}\!\left(
\frac{\omega_{2}y}{\sqrt{2E_{\star}}}
\right).
\end{equation}

Hence, the resulting toy wave function reads
\begin{equation}
\Psi_{\mathrm{toy}}(x,y)=
N\,\frac{1}{\sqrt{p_{x}(x)p_{y}(y)}}
\exp\!\left[i\big(S_{x}(x)+S_{y}(y)\big)\right],
\end{equation}
with $N=C_{x}C_{y}$. The guidance equations then reduce to
\begin{equation}
\dot{x}(t)=\sqrt{\omega_{1}^{2}x(t)^{2}+2E_{\star}},
\qquad
\dot{y}(t)=-\sqrt{\omega_{2}^{2}y(t)^{2}+2E_{\star}},
\end{equation}
which integrate to
\begin{align}
x(t)=&\,\frac{\sqrt{2E_{\star}}}{\omega_{1}}
\sinh\!\left[\omega_{1}(t-t_{0})+\eta_{0}\right],
\\
y(t)=&\,\frac{\sqrt{2E_{\star}}}{\omega_{2}}
\sinh\!\left[-\omega_{2}(t-t_{0})+\zeta_{0}\right],
\end{align}
with
\begin{equation}
\eta_{0}=\operatorname{arsinh}\!\left(
\frac{\omega_{1}x_{0}}{\sqrt{2E_{\star}}}
\right),\qquad
\zeta_{0}=\operatorname{arsinh}\!\left(
\frac{\omega_{2}y_{0}}{\sqrt{2E_{\star}}}
\right).
\end{equation}
These expressions provide explicit Bohmian trajectories in minisuperspace. 
Substituting them into
Eqs.~\eqref{eq:alpha_t} and \eqref{eq:HBohm_xy} yields a parametric 
representation of the Bohmian Hubble parameter as a function of the scale 
factor.

 \begin{figure}[!]
  \centering
  \includegraphics[width=0.5\textwidth]{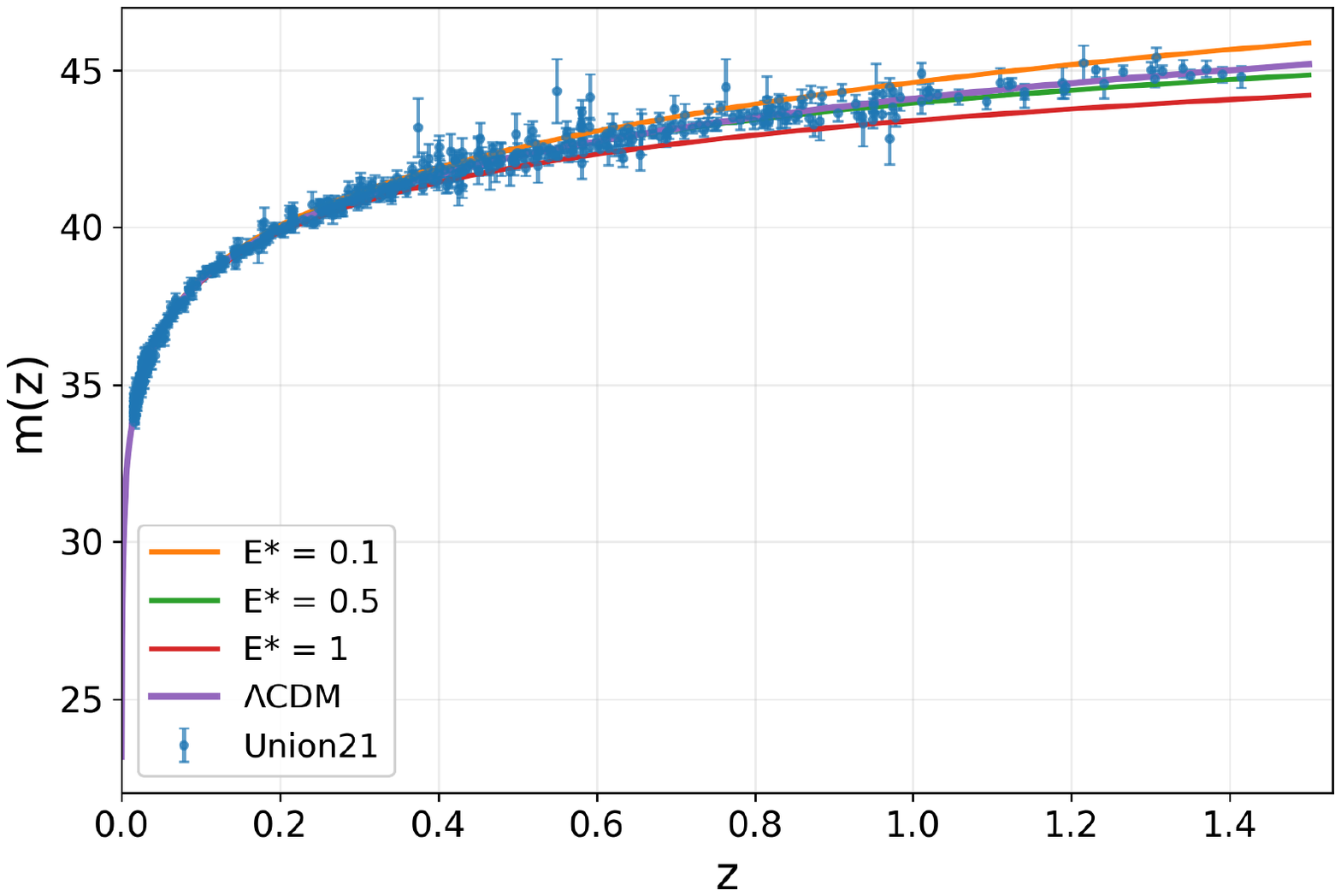}
\caption{{\it{Distance modulus $m(z)$ as a function of redshift for the Bohmian 
model,
compared with a reference flat $\Lambda$CDM cosmology and the Union21 Type Ia 
supernova compilation (points with $1\sigma$ error bars) from
\cite{Pan-STARRS1:2017jku}. The Bohmian 
background 
is generated from the toy-model parameters $w_1=1$, $w_2=\sqrt{2}$, 
$A^2=\frac{8}{3}$, $\eta_0=0$, and 
$\zeta_0=\operatorname{arsinh}\!\big(\sqrt{8/3}\big)$, and the three colored 
curves correspond to $E^\ast=0.1$, $0.5$, and $1$. For each $E^\ast$, the 
luminosity distance is computed assuming spatial flatness via 
$d_L(z)=(1+z)c\int_0^z \mathrm{d}z'/H(z')$ and converted to 
$m(z)=5\log_{10}(d_L/\mathrm{Mpc})+25$, after normalizing the model to 
$H(z{=}0)=H_0$ with $H_0=70~\mathrm{km\,s^{-1}\,Mpc^{-1}}$. The black curve is 
the flat $\Lambda$CDM prediction with 
$(\Omega_{m0},\Omega_{\Lambda0})=(0.3,0.7)$ and the same 
$H_0$.}}}
  \label{fig:mz}
\end{figure}

Fig. \ref{fig:mz} displays the redshift-distance relation in the 
observational form of the distance modulus $m(z)$. The three Bohmian-model 
curves are obtained by evolving the toy-model trajectory specified by $w_1=1$, 
$w_2=\sqrt{2}$, $A^2=\frac{8}{3}$, $\eta_0=0$, and 
$\zeta_0=\operatorname{arsinh}(\sqrt{8/3})$, and by converting the resulting 
expansion history into $d_L(z)$ through the flat-universe relation 
$d_L(z)=(1+z)c\int_0^z \mathrm{d}z'/H(z')$. The parameter $E^\ast$ controls the 
trajectory and therefore shifts the predicted $m(z)$ curve. To express 
distances in standard units, the model 
Hubble function is rescaled to satisfy $H(z{=}0)=H_0$ with 
$H_0=70~\mathrm{km\,s^{-1}\,Mpc^{-1}}$. For comparison, the black curve 
provides a conventional 
baseline given by flat $\Lambda$CDM with 
$(\Omega_{m0},\Omega_{\Lambda0})=(0.3,0.7)$, 
while the Union21 points indicate the observed trend and scatter of Type Ia 
supernova measurements over the plotted redshift range.
We stress that the Bohmian trajectories obtained here are not expectation 
values 
but represent individual cosmological histories guided by the Wheeler-DeWitt 
wave function. Different choices of the spectral weight function correspond to 
different physical boundary conditions in minisuperspace. As we observe, for  
$E^\ast$ values between $0.1$ and $0.5$ we obtain a very good agreement with 
observations.

We mention here that we do not perform a statistical fit to the supernova data, 
and the comparison is intended to show that the Bohmian expansion 
histories can closely reproduce the observed late-time behaviour while 
differing at earlier epochs. Note that since the  Bohmian 
quantum potential introduces an extra term into the effective Friedmann 
equations, the Hubble rate $ H(z)$ at late times is not determined by matter 
alone, it also depends on Bohmian parameters such as $E^\ast$. Hence, this 
dynamical framework can match both the early-universe calibration and a higher 
local measurement of the current Hubble function value $H_0$, offering an 
interesting way to alleviating the Hubble tension. Definitely the full 
confrontation with supernova data, as well as other cosmological datasets,  
will be crucial towards this direction. Since it lies 
beyond the scope of the 
present work, it will be performed elsewhere.

 \begin{figure}[!]
  \centering
  \includegraphics[width=0.49\textwidth]{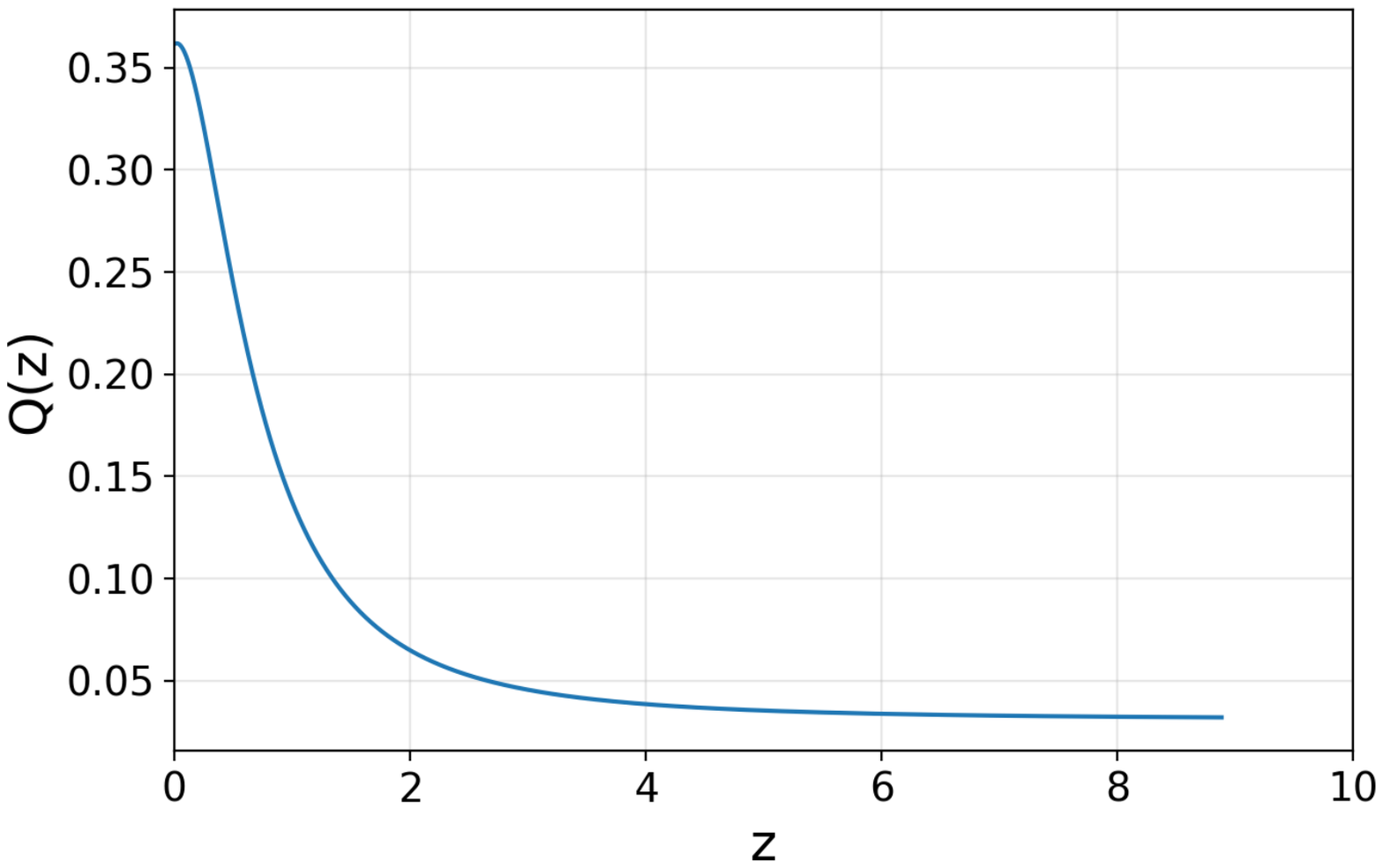}
\caption{{\it{The Bohmian quantum potential $Q(z)$ along the toy-model Bohmian 
trajectory, expressed as a function of redshift $z=1/a-1$, on the expanding 
(past) branch $z\ge 0$. The trajectory is generated with $w_1=1$, 
$w_2=\sqrt{2}$, $A^2=\frac{8}{3}$, $\eta_0=0$, and 
$\zeta_0=\operatorname{arsinh}\!\big(\sqrt{8/3}\big)$, and the curve shown 
corresponds to $E^\ast=1$ (representative case).}}}
\label{fig:Qz}
\end{figure}
In Fig. \ref{fig:Qz} we show how the Bohmian quantum potential varies with 
redshift when evaluated along the model trajectory. Using the toy-model 
constants $w_1=1$, $w_2=\sqrt{2}$, $A^2=\frac{8}{3}$, $\eta_0=0$, and 
$\zeta_0=\operatorname{arsinh}(\sqrt{8/3})$, we compute $Q$ along the Bohmian 
solution and reparameterize it by redshift using $z=1/a-1$, restricting to the 
expanding (past) branch with $z\ge 0$. The plotted curve corresponds to 
$E^\ast=1$ and illustrates that $Q(z)$ is larger at high redshift (small scale 
factor) and decreases toward low redshift, indicating that the quantum 
contribution is most pronounced in the early-time regime and becomes less 
important as the universe expands.
Although the quantum potential increases as $z\to0$, its contribution to the
guidance equations becomes negligible compared to the classical terms, ensuring
the recovery of the late-time classical cosmological behaviour. Hence, this 
behaviour explicitly illustrates the Bohmian mechanism for the recovery of 
classical cosmology, with quantum effects dominating at early times and 
becoming negligible as the universe expands.

 \begin{figure}[ht]
  \centering
  \includegraphics[width=0.48\textwidth]{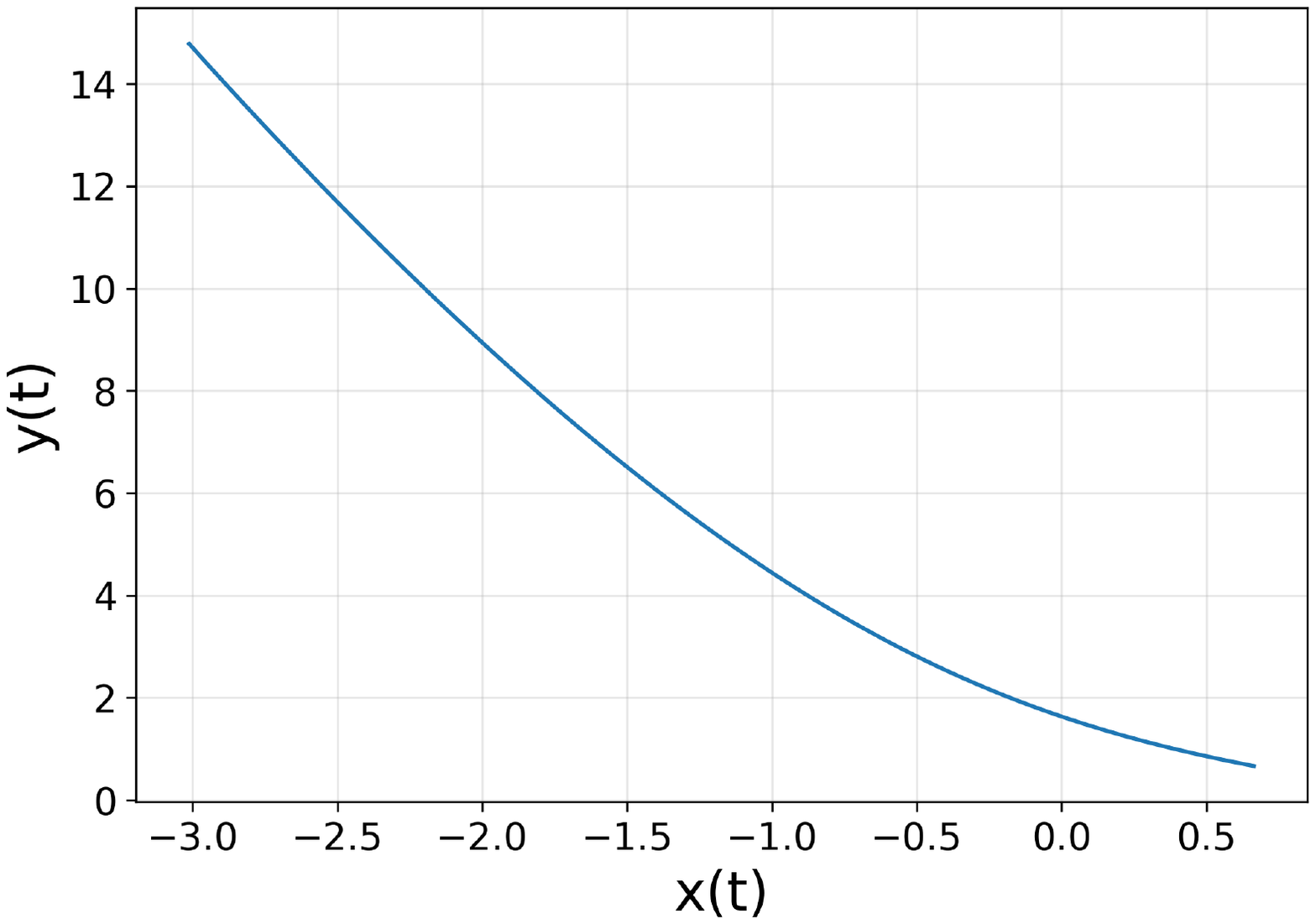}
\caption{{\it{Parametric Bohmian minisuperspace trajectory $(x(t),y(t))$ for 
the Wheeler-DeWitt toy mode $\Psi_{\rm toy}$ obtained in the WKB regime.
The curve represents a single deterministic cosmological history rather than an 
expectation value, and follows from integrating the guidance equations, 
yielding the analytic solutions
$x(t)=\frac{\sqrt{2E_\star}}{\omega_1}\sinh\!\big[\omega_1(t-t_0)+\eta_0\big]$ 
and
$y(t)=\frac{\sqrt{2E_\star}}{\omega_2}\sinh\!\big[-\omega_2(t-t_0)+\zeta_0\big]
$.
In the plot we use $E_\star=1$, $\omega_1=1$, $\omega_2=\sqrt{2}$, $\eta_0=0$, 
and $\zeta_0=\operatorname{arsinh}\!\big(\sqrt{8/3}\big)$. The trajectory lies 
in the physical domain $y^2-x^2>0$, which corresponds to a positive scale 
factor.}}}
\label{fig:xytraj}
\end{figure}

Finally, in Fig. \ref{fig:xytraj} we depict  a representative Bohmian 
trajectory in the 
two-dimensional minisuperspace spanned by the canonical variables $(x,y)$, 
which are related to the Friedmann-Robertson-Walker scale factor $\alpha(t)$ 
and the 
homogeneous scalar field $\phi(t)$ via the canonical transformation
$x=A\,\alpha^{3/2}\sinh(c\phi)$ and $y=A\,\alpha^{3/2}\cosh(c\phi)$.
This implies the constraint $y^2-x^2=A^2\alpha^3$, so the physically admissible 
region corresponds to $y^2-x^2>0$ (i.e.\ $\alpha>0$). For the sharply peaked 
toy wave function $\Psi_{\rm toy}=R\,e^{iS}$, the Bohmian guidance equations 
reduce to a pair of decoupled first-order equations for $x(t)$ and $y(t)$, 
which integrate to the explicit hyperbolic-sine parametric form shown in the 
caption. Each curve therefore represents a single deterministic cosmological 
history guided by the pilot-wave phase $S(x,y)$, from which one may reconstruct 
$\alpha(t)$ (and hence the expansion rate) along the trajectory.

\section{Conclusions}
\label{Conclusions}

Quantum cosmology provides a natural framework for exploring the interplay 
between quantum theory and the large-scale dynamics of the universe, 
particularly in regimes where classical general relativity is expected to break 
down. Within this context, minisuperspace models offer a tractable yet 
physically meaningful setting in which the Wheeler-DeWitt equation can be 
formulated and solved explicitly. However, extracting a notion of cosmological 
evolution from a timeless quantum constraint equation requires an 
interpretational framework capable of assigning physical meaning to the wave 
function of the universe. The de Broglie-Bohm (pilot-wave) formulation is 
particularly well suited for this purpose, as it allows one to define definite 
trajectories in configuration space without invoking an external time or 
measurement process.

In this work we have developed a Bohmian quantum cosmological model for a 
spatially flat Friedmann-Robertson-Walker universe containing a single scalar 
field 
whose potential unifies the description of cold dark matter and dark energy at 
the background level. Starting from the Einstein-Hilbert action supplemented by 
a scalar field, we constructed the minisuperspace Lagrangian and its canonical 
Hamiltonian formulation. A key step in our analysis was the identification of a 
nontrivial canonical transformation that maps the minisuperspace dynamics into 
that of a two-dimensional hyperbolic oscillator with a fixed relation between 
its frequencies. This reformulation renders the Wheeler-DeWitt equation 
particularly transparent and allows for its exact solution by separation of 
variables in terms of parabolic cylinder functions.

A central result of our analysis is that the Wheeler-DeWitt equation admits a 
continuous family of solutions labelled by a separation constant, rather than a 
discrete spectrum of harmonic-oscillator-like quantum numbers. This feature 
reflects the constrained, Klein-Gordon-type nature of the Wheeler-DeWitt 
equation and the absence of a standard Schrödinger time parameter and 
associated square-integrability condition. Within the Bohmian framework, we 
decomposed the wave function into amplitude and phase, derived the 
corresponding quantum Hamilton-Jacobi equation and quantum potential, and 
obtained deterministic guidance equations for the minisuperspace variables. 
From 
these trajectories we constructed a well-defined Bohmian Hubble parameter, 
directly expressed in terms of the pilot-wave phase. By means of a 
Wheeler-DeWitt-derived toy wave function with a sharply peaked spectrum, we 
explicitly demonstrated how analytic Bohmian trajectories and a concrete 
cosmological expansion history can emerge from the underlying quantum 
description, closely reproducing the late-time $\Lambda$CDM behaviour while 
exhibiting quantum modifications at earlier epochs.

The framework developed here opens several directions for further 
investigation. Natural extensions include the incorporation of spatial 
curvature, additional scalar degrees of freedom, or perturbations around the 
homogeneous background, which would allow one to explore quantum effects beyond 
the minisuperspace approximation. It would also be of interest to study how 
different choices of boundary conditions in minisuperspace, encoded in the 
spectral weight function, influence the Bohmian trajectories and the resulting 
cosmic histories. Ultimately, confronting Bohmian quantum cosmological models 
with observational data from Cosmic Microwave Background (CMB), Baryonic
Acoustic Oscillations (BAO), power spectra and large-scale structure 
observations, may help assess whether pilot-wave dynamics can provide 
a viable and distinctive description of the quantum origin and evolution of the 
universe, complementing more conventional approaches to quantum cosmology.

\begin{acknowledgments}
The authors gratefully acknowledges  the 
contribution of 
the LISA Cosmology Working Group (CosWG), as well as support from the COST 
Actions CA21136 -  Addressing observational tensions in cosmology with 
systematics and fundamental physics (CosmoVerse)  - CA23130, Bridging 
high and low energies in search of quantum gravity (BridgeQG)  and CA21106 -  
 COSMIC WISPers in the Dark Universe: Theory, astrophysics and 
experiments (CosmicWISPers). 

\end{acknowledgments}

\bibliographystyle{unsrt}

\begin{thebibliography}{99}

\bibitem{DeWitt1967}
B.~S.~DeWitt,
  {\it{Quantum Theory of Gravity. I. The Canonical Theory}},
Phys.\ Rev.\ {\bf 160}, 1113 (1967).

 \bibitem{Wheeler:1968Superspace}
  J.~A.~Wheeler,
  {\it{Superspace and the Nature of Quantum Geometrodynamics}},
  in {\it{Battelle Rencontres: 1967 Lectures in Mathematics and Physics},
  ed.\ C.~M.~DeWitt and J.~A.~Wheeler}, pp.\ 242-307 (1968).

 
\bibitem{Hartle:1983ai}
J.~B.~Hartle and S.~W.~Hawking,
 {\it{Wave Function of the Universe,}}
Phys. Rev. D \textbf{28}, 2960-2975 (1983)
  [\href{https://doi.org/10.1103/PhysRevD.28.2960}{{\tt 
doi:10.1103/10.1103/PhysRevD.28.2960}}].

 
  

\bibitem{Page1984}
  D.~N.~Page,
  {\it{Classical stability of round and squashed seven-spheres in 
eleven-dimensional supergravity}},
  Phys.\ Rev.\ D {\bf 28}, 2976-2982 (1983)
  [\href{https://doi.org/10.1103/PhysRevD.28.2976}{{\tt 
doi:10.1103/PhysRevD.28.2976}}].


  

  
  

\bibitem{Kiefer1988}
  C.~Kiefer,
  {\it{Wave packets in minisuperspace}},
  Phys.\ Rev.\ D {\bf 38}, 1761-1772 (1988)
  [\href{https://doi.org/10.1103/PhysRevD.38.1761}{{\tt 
doi:10.1103/PhysRevD.38.1761}}].

\bibitem{Halliwell:1988Wheeler}
J.~J.~Halliwell,
 {\it{Derivation of the Wheeler-DeWitt equation from a path integral for 
minisuperspace models,}}
Phys. Rev. D \textbf{38}, 2468 (1988)
  [\href{https://doi.org/10.1103/PhysRevD.38.2468}{{\tt 
doi:10.1103/PhysRevD.38.2468}}].

 


\bibitem{Falciano2012}
  F.~T.~Falciano, R.~Pereira, N.~Pinto-Neto and E.~S.~Santini,
  {\it{The Wheeler-DeWitt Quantization Can Solve the Singularity Problem}},
  Phys.\ Rev.\ D {\bf 86}, 063504 (2012)
  [\href{https://arxiv.org/abs/1206.4021}{{\tt arXiv:1206.4021}}].

  
\bibitem{Kiefer:2008Cosmology}
C.~Kiefer and B.~Sandhoefer,
{\it{Quantum cosmology}},
Classical and Quantum Gravity \textbf{21} (2008)
  [\href{https://doi.org/10.1515/zna-2021-0384}{{\tt 
doi:10.1515/zna-2021-0384}}].

 



\bibitem{Vilenkin:1984Creation}
  A.~Vilenkin,
  {\it{Quantum creation of universes}},
  Phys.\ Rev.\ D {\bf 30}, 509-511 (1984).

\bibitem{Linde:1984InflationaryCreation}
  A.~D.~Linde,
  {\it{Quantum creation of the inflationary universe}},
  Lett.\ Nuovo Cim.\ {\bf 39}, 401-405 (1984).

  
  
  
\bibitem{Misner:1969hg}
  C.~W.~Misner,
  {\it{Mixmaster universe}},
  Phys.\ Rev.\ Lett.\ {\bf 22}, 1071-1074 (1969)
  [\href{https://doi.org/10.1103/PhysRevLett.22.1071}{{\tt 
doi:10.1103/PhysRevLett.22.1071}}].

\bibitem{HawkingWu}
  S.~W.~Hawking and Z.~C.~Wu,
  {\it{Numerical calculations of minisuperspace cosmological models}},
  Phys.\ Lett.\ B {\bf 151}, 15-20 (1985)
  [\href{https://doi.org/10.1016/0370-2693(85)90815-9}{{\tt 
doi:10.1016/0370-2693(85)90815-9}}].
 
 
\bibitem{Kuchar:1989tj}
  K.~V.~Kucha\v{r} and M.~P.~Ryan,
  {\it{Is minisuperspace quantization valid?: Taub in mixmaster}},
  Phys.\ Rev.\ D {\bf 40}, 3982-3996 (1989)
  [\href{https://doi.org/10.1103/PhysRevD.40.3982}{{\tt 
doi:10.1103/PhysRevD.40.3982}}].


  
 

\bibitem{Vilenkin:1994rn}
  A.~Vilenkin,
  {\it{Approaches to quantum cosmology}},
  Phys.\ Rev.\ D {\bf 50}, 2581-2594 (1994)
  [\href{https://arxiv.org/abs/gr-qc/9403010}{{\tt arXiv:gr-qc/9403010}}].

 


\bibitem{Giulini:1994dx}
  D.~Giulini and C.~Kiefer,
  {\it{Wheeler-DeWitt metric and the attractivity of gravity}},
  Phys.\ Lett.\ A {\bf 193}, 21-24 (1994)
  [\href{https://arxiv.org/abs/gr-qc/9405040}{{\tt arXiv:gr-qc/9405040}}].


\bibitem{Gousheh:2000au}
  S.~S.~Gousheh and H.~R.~Sepangi,
  {\it{Wave packets and initial conditions in quantum cosmology}},
  Phys.\ Lett.\ A {\bf 272}, 304-312 (2000)
  [\href{https://arxiv.org/abs/gr-qc/0006094}{{\tt arXiv:gr-qc/0006094}}].

\bibitem{Garattini:2014rwa}
  R.~Garattini and E.~N.~Saridakis,
  {\it{Gravity's Rainbow: a bridge towards Ho\v{r}ava-Lifshitz gravity}},
  Eur.\ Phys.\ J.\ C {\bf 75}, 343 (2015)
  [\href{https://arxiv.org/abs/1411.7257}{{\tt arXiv:1411.7257}}].

\bibitem{Kiefer:2005AnnPhysReview}
  C.~Kiefer,
  {\it{Quantum Gravity: General Introduction and Recent Developments}},
  Annalen Phys.\ {\bf 15}, 129-148 (2005)
  [\href{https://arxiv.org/abs/gr-qc/0508120}{{\tt arXiv:gr-qc/0508120}}].
  
\bibitem{Scherrer2004}
  R.~J.~Scherrer,
  {\it{Purely kinetic k-essence as unified dark matter}},
  Phys.\ Rev.\ Lett.\ {\bf 93}, 011301 (2004)
  [\href{https://arxiv.org/abs/astro-ph/0402316}{{\tt arXiv:astro-ph/0402316}}].

\bibitem{Bertacca2007}
  D.~Bertacca, S.~Matarrese and M.~Pietroni,
  {\it{Unified dark matter in scalar field cosmologies}},
  Mod.\ Phys.\ Lett.\ A {\bf 22}, 2893-2907 (2007)
  [\href{https://arxiv.org/abs/astro-ph/0703259}{{\tt arXiv:astro-ph/0703259}}].

\bibitem{Arbey:2006it}
  A.~Arbey,
  {\it{Dark fluid: A complex scalar field to unify dark energy and dark 
matter}},
  Phys.\ Rev.\ D {\bf 74}, 043516 (2006)
  [\href{https://arxiv.org/abs/astro-ph/0601274}{{\tt arXiv:astro-ph/0601274}}].

\bibitem{Kouniatalis:2025qfz}
  G.~Kouniatalis,
  {\it{The Wave Function of the Universe and Inflation}},
  (2025)
  [\href{https://arxiv.org/abs/2510.04775}{{\tt arXiv:2510.04775}}].

\bibitem{Lim:2010yk}
  E.~A.~Lim, I.~Sawicki and A.~Vikman,
  {\it{Dust of Dark Energy}},
  JCAP {\bf 05}, 012 (2010)
  [\href{https://arxiv.org/abs/1003.5751}{{\tt arXiv:1003.5751}}].

  \bibitem{Kouniatalis:2025orn}
  G.~Kouniatalis and E.~N.~Saridakis,
  {\it{Inflation from a generalized exponential plateau: towards extra 
suppressed tensor-to-scalar ratios}},
  JCAP {\bf 11}, 038 (2025)
  [\href{https://arxiv.org/abs/2507.17721}{{\tt arXiv:2507.17721}}].

\bibitem{Piattella:2009kt}
  O.~F.~Piattella, D.~Bertacca, M.~Bruni and D.~Pietrobon,
  {\it{Unified Dark Matter models with fast transition}},
  JCAP {\bf 01}, 014 (2010)
  [\href{https://arxiv.org/abs/0911.2664}{{\tt arXiv:0911.2664}}].

\bibitem{Koutsoumbas:2017fxp}
  G.~Koutsoumbas, K.~Ntrekis, E.~Papantonopoulos and E.~N.~Saridakis,
  {\it{Unification of Dark Matter-Dark Energy in Generalized Galileon 
Theories}},
  JCAP {\bf 02}, 003 (2018)
  [\href{https://arxiv.org/abs/1704.08640}{{\tt arXiv:1704.08640}}].

  \bibitem{ColinPintoNeto:2017MatterBounceDE}
  S.~Colin and N.~Pinto-Neto,
  {\it{Quantum matter bounce with a dark energy expanding phase}},
  Phys.\ Rev.\ D {\bf 96}, 063502 (2017)
  [\href{https://arxiv.org/abs/1706.03037}{{\tt arXiv:1706.03037}}].

\bibitem{BacalhauPintoNetoVitenti:2018Spectra}
  A.~P.~Bacalhau, N.~Pinto-Neto and S.~D.~P.~Vitenti,
  {\it{Consistent Scalar and Tensor Perturbation Power Spectra in Single Fluid 
Matter Bounce with Dark Energy Era}},
  Phys.\ Rev.\ D {\bf 97}, 083517 (2018)
  [\href{https://arxiv.org/abs/1706.08830}{{\tt arXiv:1706.08830}}].
  
 


\bibitem{Leon:2022oyy}
  G.~Leon, A.~Paliathanasis, E.~N.~Saridakis and S.~Basilakos,
  {\it{Unified dark sectors in scalar-torsion theories of gravity}},
  Phys.\ Rev.\ D {\bf 106}, 024055 (2022)
  [\href{https://arxiv.org/abs/2203.14866}{{\tt arXiv:2203.14866}}].



 

\bibitem{VargasMoniz:2010zz}
  P.~Vargas Moniz,
  {\it{Quantum cosmology - the supersymmetric perspective: Vol.\ 1: 
Fundamentals}},
  Lect.\ Notes Phys.\ {\bf 803}, 1-351 (2010)
  [\href{https://doi.org/10.1007/978-3-642-11575-2}{{\tt 
doi:10.1007/978-3-642-11575-2}}].

  
  

\bibitem{deBroglie1926}
  L.~de~Broglie,
  {\it{Interference and Corpuscular Light}},
  Nature {\bf 118}, 441-442 (1926)
  [\href{https://doi.org/10.1038/118441b0}{{\tt doi:10.1038/118441b0}}].

\bibitem{Bohm1952}
  D.~Bohm,
  {\it{A Suggested Interpretation of the Quantum Theory in Terms of `Hidden' 
Variables. I}},
  Phys.\ Rev.\ {\bf 85}, 166-179 (1952)
  [\href{https://doi.org/10.1103/PhysRev.85.166}{{\tt 
doi:10.1103/PhysRev.85.166}}].

\bibitem{BohmHiley1993}
  D.~Bohm and B.~J.~Hiley,
  {\it{The Undivided Universe}},
  Routledge (1993).

\bibitem{Holland1993}
  P.~R.~Holland,
  {\it{The Quantum Theory of Motion}},
  Cambridge University Press (1993).

\bibitem{PintoNeto:2004BohmQC}
  N.~Pinto-Neto,
  {\it{The Bohm Interpretation of Quantum Cosmology}},
  Found.\ Phys.\  {\bf 35}, 577-603 (2005)
  [\href{https://arxiv.org/abs/gr-qc/0410117}{{\tt arXiv:gr-qc/0410117}}].
  
  

\bibitem{PintoNeto:2021dBBQCReview}
  N.~Pinto-Neto,
  {\it{The de Broglie-Bohm Quantum Theory and its Application to Quantum 
Cosmology}},
  Universe {\bf 7}, 134 (2021)
  [\href{https://arxiv.org/abs/2111.03057}{{\tt arXiv:2111.03057}}].

\bibitem{PintoNetoStruyve:2019BohmQGCosmo}
  N.~Pinto-Neto and W.~Struyve,
  {\it{Bohmian quantum gravity and cosmology}},
  in {\it{Applied Bohmian Mechanics: From Nanoscale Systems to Cosmology}} (2nd 
ed.),
  eds.\ X.~Oriols Pladevall and J.~Mompart (Jenny Stanford Publishing, 2019)
  [\href{https://arxiv.org/abs/1801.03353}{{\tt arXiv:1801.03353}}].

  

  
  
  
\bibitem{Shojai2005}
  A.~Shojai and M.~Shirinifard,
  {\it{Bohmian quantum cosmology}},
  Int.\ J.\ Mod.\ Phys.\ D {\bf 14}, 1333-1345 (2005)
  [\href{https://arxiv.org/abs/gr-qc/0504138}{{\tt arXiv:gr-qc/0504138}}].

\bibitem{Colistete:1997sf}
  R.~Colistete, Jr., J.~C.~Fabris and N.~Pinto-Neto,
  {\it{Singularities and classical limit in quantum cosmology with scalar 
fields}},
  Phys.\ Rev.\ D {\bf 57}, 4707-4717 (1998)
  [\href{https://arxiv.org/abs/gr-qc/9711047}{{\tt arXiv:gr-qc/9711047}}].

  


  

\bibitem{PintoNetoFabris:2013dBBQC}
  N.~Pinto-Neto and J.~C.~Fabris,
  {\it{Quantum cosmology from the de Broglie-Bohm perspective}},
  Class.\ Quant.\ Grav.\  {\bf 30}, 143001 (2013)
  [\href{https://arxiv.org/abs/1306.0820}{{\tt arXiv:1306.0820}}].
  


 

\bibitem{FalcianoPintoNetoSantini:2007InflationaryNS}
  F.~T.~Falciano, N.~Pinto-Neto and E.~S.~Santini,
  {\it{An Inflationary Non-singular Quantum Cosmological Model}},
  Phys.\ Rev.\ D {\bf 76}, 083521 (2007)
  [\href{https://arxiv.org/abs/0707.1088}{{\tt arXiv:0707.1088}}].

  
  
\bibitem{Paliathanasis:2014iva}
A.~Paliathanasis, S.~Basilakos, E.~N.~Saridakis, S.~Capozziello, K.~Atazadeh, 
F.~Darabi and M.~Tsamparlis,
  {\it{New Schwarzschild-like solutions in f(T) gravity through Noether 
symmetries,}}
Phys. Rev. D \textbf{89}, 104042 (2014)
  [\href{https://arxiv.org/abs/1402.5935}{{\tt arXiv:1402.5935}}].

 

\bibitem{Paliathanasis:2015aos}
A.~Paliathanasis,
  {\it{$f(R)$-gravity from Killing Tensors,}}
Class. Quant. Grav. \textbf{33}, no.7, 075012 (2016)
  [\href{https://arxiv.org/abs/1512.03239}{{\tt arXiv:1512.03239}}].

  
\bibitem{Dimakis:2016mip}
N.~Dimakis, A.~Karagiorgos, A.~Zampeli, A.~Paliathanasis, T.~Christodoulakis 
and 
P.~A.~Terzis,
  {\it{General Analytic Solutions of Scalar Field Cosmology with Arbitrary 
Potential,}}
Phys. Rev. D \textbf{93}, no.12, 123518 (2016)
  [\href{https://arxiv.org/abs/1604.05168}{{\tt arXiv:1604.05168}}].

  
\bibitem{DemaerelStruyve:2019Orderings}
  T.~Demaerel and W.~Struyve,
  {\it{Elimination of cosmological singularities in quantum cosmology by 
suitable operator orderings}},
  Phys.\ Rev.\ D {\bf 100}, 046008 (2019)
  [\href{https://arxiv.org/abs/1904.09244}{{\tt arXiv:1904.09244}}].
  
  


\bibitem{KieferPeter:2022TimeQC}
  C.~Kiefer and P.~Peter,
  {\it{Time in quantum cosmology}},
  Universe {\bf 8}, 36 (2022)
  [\href{https://arxiv.org/abs/2112.05788}{{\tt arXiv:2112.05788}}].
  
\bibitem{Pan-STARRS1:2017jku}
D.~M.~Scolnic \textit{et al.} [Pan-STARRS1],
  {\it{The Complete Light-curve Sample of Spectroscopically Confirmed SNe Ia 
from Pan-STARRS1 and Cosmological Constraints from the Combined Pantheon 
Sample,}}
Astrophys. J. \textbf{859}, no.2, 101 (2018)
  [\href{https://arxiv.org/abs/1710.00845}{{\tt arXiv:1710.00845}}].

\bibitem{Olver:2010:NHMF}
F.~W.~J.~Olver, D.~W.~Lozier, R.~F.~Boisvert and C.~W.~Clark,
{\it{NIST Handbook of Mathematical Functions,}}
Cambridge University Press, New York (2010)
[\href{https://doi.org/10.5555/1830479}{{\tt doi:10.5555/1830479}}].

 

 


\end{thebibliography}

\end{document}